\documentclass[11pt,twoside]{article}
\usepackage{asp2006}
\usepackage{graphics}
\usepackage{lscape}
\def\edcomment#1{\iffalse\marginpar{\raggedright\sl#1\/}\else\relax\fi}
\reversemarginpar

\begin{document}
\title{Some new reflections on Mr. Palomar}
\author{Orfeu Bertolami}
\affil{Instituto Superior T\'ecnico, Departamento de F\'\i sica, \\
Av. Rovisco Pais, 1049-001 Lisboa, Portugal \\
and \\
Instituto de Plasmas e Fus\~ao Nuclear, Instituto Superior T\'ecnico,
Lisboa.}

\begin{abstract}The character Mr. Palomar, the alter-ego of the Italian author
Italo Calvino,
appeared for the first time in 1975 on the pages of the ``Il Corriere della 
Sera'',
and then more or less regularly till its debut as a book in 1983. Through
illuminating thoughts and reflections based on observations, for instance, of
sea waves, Mr. Palomar discovers that they
induce a peaceful and inspirational state of mind that prevents
coronary and mental illnesses, and also holds the key to capturing the
complexity of the world reducing it into its most elementary mechanisms.
In this contribution I will survey some of Mr. Palomar's thoughts while he
observes the sky and speculate on others that he might have explored if he
shared our contemporary knowledge of the
cosmos.
I will also discuss the thoughts of other authors on how,
cosmological thinking affects the human condition.
\end{abstract}

\vspace{-0.5cm}

\section{Cosmology and ``cosmo-vision''}
\label{sec:cosmology}

An impressive body of experimental and observational facts, models,
conjectures and hypotheses give shape to our current cosmo-vision --
``Weltanschauung''. This evidence arises from a wide variety of sources including
astronomy, physics, chemistry,
geology, paleontology, biology, genetics, archeology and history,
providing a most exciting and vivid picture of humankind and the Universe's evolution.
Most likely, this laborious
work of gathering ideas and observation supporting this cosmo-vision,
which took various generations
of researchers many years and that will likely
last many more in the foreseeable future, is one of the highest achievements of human endeavour. 
It allows for the understanding of the articulation of a variety of ``origins'',
from the origin of the Universe to the emergence of life on Earth and its subsequent
evolution, to the first human social formations (see \cite{Bertolami06a} for a discussion).

Studies of the Cosmic Microwave Background Radiation (CMBR) and theoretical cosmology allow
estimating the age of the Universe as about
13.7 thousand million years (Gys). The CMBR corresponds to the surface of last
scattering, when the Universe turned transparent to electromagnetic radiation (and 
the approximate time of transition from the early dominance by the
density energy of radiation to the dominance of matter), about 370 thousands
years after the Big Bang. The first galaxies and stars were formed about
10 Gys ago. Radiative dating allows estimating the age of Earth and Solar System as about
4.5 Gys. Evidence arising from dating of stromatolite's fossils suggests
that life on Earth appeared 3.6 Gys ago. The first macroscopic organisms
first appeared about 700 million years ago. The tectonic sea-floor spreading
which resulted in the Atlantic ocean took place about 100 million
years ago. Our primate ancestors first walked on two legs about
3 million years ago. The first human socio-cultural formations appeared
40 thousand years ago and the first human artistic
imagery, impressive cave and rock paintings and rock art found in Europe and Australia, date
back to 15 to 45 thousand years.

It is quite evident that technological developments have also shaped and conditioned the
evolution of humankind, most notably from the XIX century onwards. Most likely,
this development will be even more conspicuous in the future.
Nevertheless, however special it might be (see contributions by Joel Primack and Nancy Abrams),
the impact that the discovery of our modest standing
in the Universe is much less apparent in our culture. Still, strong arguments can be put forward
that science, and cosmology in particular,
do permeate our culture. This can be observed in
contemporary art,
science fiction, cinema\footnote{The scene of a boy in
one of Woody Allen's films refusing to eat because the Universe was expanding is one that
strikes quite vividly
my memory.}, music and so on. A ``visual poetry" inspired by a cosmic
view can be recognized in many ancient cultures (see intervention of Chanda Carey).
Literature, specially contemporary literature, is receptive to the appeal of cosmology.
I will concentrate on some examples I find
particularly representative.

Pessimism about our cosmic condition and about the emergence of life can be
found in the writings of important authors. In one of his well known essays,
Bertrand Russell (1872 -- 1970) is quite emphatic about the influence that the knowledge
of catastrophic cosmic
events and the inevitable death sentence on the Universe and human achievements as a result of 
the Second Law of Thermodynamics \citep{Russell}:

\noindent
{\it `` ... all the inspiration, all the noonday brightness of the human
genius are destined to extinction in the vast death of the solar system,
and the whole temper of Man's achievement must inevitably be buried
beneath the debris of a universe.''}

Rather against the current anthropic reasoning, which argues in favour
of the near inevitability of the emergence of life
in the Universe in light of examples like the recent discovery of a profusion extra-solar
planets\footnote{An infinite number of worlds was defended
by Giordano Bruno (1548 -- 1600), as the only way to God exert his omnipotence \citep{Bruno}.
For this heretical view he was burnt at the
stake at the Market of Flowers in Rome in 1600.}, Nobel laureate, French biologist,
Jacques Monod (1910 -- 1976) expresses his
utter pessimism \citep{Monod}:

\noindent
{\it ``... The Universe was not pregnant with life nor biosphere with
man. Our number came up in the Monte Carlo game ...}

\noindent
{\it then man must at last
wake out of his millinery dream, and in doing so wake to his total solitude,
his fundamental isolation ...}

\noindent
{\it Man knows at last that he is alone in the unfeeling immensity of
the universe, out of which he has emerged only by chance.''}

In a contrasting tone, I find, no author has discussed the influence of the physical world in such
an optimistic and rather analytic way than the Italian author Italo Calvino (1923 -- 1985).
The character in question is Mr. Palomar\footnote{Other writings of Calvino on the Universe, 
include ``Le cosmocomiche'' (1965),
``Ti con zero'' (1967), ``La memoria del mondo e altre storie cosmocomiche'' (1968).},
after the mountain where the famous Hale
telescope in California enabled remarkable discoveries about the Universe from the late 1940s 
until 1980s.
Mr. Palomar, in a broad sense, the alter-ego of the Italian author himself,
appeared for the first time in 1975 on the pages of the ``Il Corriere della Sera'',
and then somewhat regularly until appearing as the central figure in a book 
as simply ``Palomar'' \citep{Calvino}.

Mr. Palomar thoughts
and reflections are quite original and illuminating, arising through observing the natural world (physical and
biological), as well as when sensing the sociological difficulties of our day to day life.
Actually, it is rather unique the way he concludes that human relations
should necessarily mirror the Universe in order to improve.

In the chapter, ``Palomar sulla spiaggia'' and, in particular, in the subsection ``Lettura di un'onda''
(Palomar on the beach - The reading of a wave)", the character discovers,
for instance, that the
observation of the waves in the sea, not only
induces a peaceful and inspirational state of mind,
but also that it holds the key to capturing the
complexity of the world reducing it into its most elementary mechanisms.

\noindent
{\it `` ... One cannot observe a wave without taking into account the complex aspects
(velocity, shape, direction) that conspire to form it.
The factors are always changing so that every wave is different than any
other, but it is also true that they are equal to each other ...''}

In the chapter Palomar in the garden, (``Palomar in giardino''), and in the subsection on the loving making of the
tortoises, (``Gli amori delle tartarughe''), the character reflects on the extraordinary forces of biological attraction
and the complication it represents for certain species, including our.

In Palomar looks at to the sky, (``Palomar guarda il cielo''), the character 
draws his conclusions when observing the moon
in the afternoon, the motion of the planets and the shinning of the stars
(``Luna di pomeriggio'', ``L'occhio e i pianeti'', ``La contemplazione delle stelle''). 
These conclusions are not always correct from the physics point of view, but nevertheless 
original and insightful.
Actually, he remarks that the contemplation of the stars requires a great effort, as one should be properly equipped with a
telescope, a chart of the constellations, a lamp, etc. He refers to the Milky Way 
as ``the terrible shinning silvered cloud''. His disquiet about the distance separating the objects in the sky is
evident, as they are beyond our understanding. Moreover, he claims that the observation of the stars
induces unstable and contradictory feelings as it suggests a too complex relationship between harmony and evolution.
And rather remarkably:

\noindent
{\it `` The uncertainty about the distance of the luminous bodies leads one to trust only on the dark!
What could be more stable than nothing?''}

I think that Calvino would be quite pleased to hear that recent discoveries suggest that the real dynamics
of the cosmos is actually ruled by dark entities, dark energy, on the largest scales, and dark matter,
on galactic and galaxy cluster's scales. The simplest candidate for dark energy, and for some the
most natural, is actually the vacuum energy density, ``nothing'' if one
assumes that the ``void'' refers to the absence of matter that does not manifest itself in the
electromagnetic spectrum, or that cannot be observed through, for instance, neutrino detectors.

In the chapter Universe as mirror (``L'Universo come specchio''), Palomar calls for a ``cosmological
thinking'' in order to improve our lives. His point is that his suffering when weighting the difficulties
in his relationships with others, would improve if his relationship with the
Universe were closer. He argues that despite the infinite
combinations, permutations and chain of consequences, all events in the Universe remain and
that should be the basic underlying feature of human relationships! Palomar goes as far as, to quote as an example,
the explosion of a supernova at the Magellanic Cloud as an event that took place long ago, but that
is still there, to be seen, admired and discussed.
One can see that as an incredible premonition of the SNe 1987a event, precisely at the Magellanic Cloud!

Let us now turn to another author, the Argentinian Ernesto S\'abato (1911).
The expansion of the Universe is one of the central issues of discussion in the ``Uno y el Universo''
 (One and the Universe) \citep{Sabato}.

Being a physicist by training, the discovery of the expansion of the Universe could not fail to impress
him. He was aware of the pioneering work of Einstein, who in 1917 defended the idea of a static Universe, and of
his Dutch colleague Willem de Sitter, who in the same year showed that a Universe dominated by the cosmological term introduced by
Einstein to keep the Universe at bay, actually, would not do the job!
He was also aware of the 1922 evolving solution of the Russian engineer Alexander
Friedmann and the subsequent discussion and solutions by the Belgium priest George Lema\^\i tre
from 1924 onwards\footnote{See \cite{Lemaitre} for a thorough discussion of the
historical developments which led to the idea of the ``day without yesterday'', the Big Bang.}.

Suspicious of purely theoretical constructions to explain the expansion of the Universe, as suggested
for instance by the British astronomer Arthur Eddington \citep{Eddington}, he expresses
also reservations about purely empirical evidence.

These concerns are ingeniously exemplified through an analogy with ichthyology. The example goes as follows:
through the repeating process of collecting fish
with a catch with a spacing of 5 cm an ichthyologist acquires knowledge that he/she expresses
in terms of two laws:

 \noindent
1) There are no fish smaller than 5 cm long.

 \noindent
2)  All fish have gills.

\noindent
From a strictly scientific point of view, any natural scientist would argue that fish belong to the
physical world, that a fellow ichthyologist is a well intentioned and competent scientist, and the catch
is the cognition apparatus. However, from the point of view of a skeptic, the first law is just a
consequence of the net employed and hence the validity of the second law might be in question.

To this criticism, a hard core ichthyologist would counter argue that fish that cannot be caught
with the available net are beyond ichthyology's
knowledge. They belong to metaphysics. Science is built upon observable entities.
Of course, on purely epistemological terms one might argue that the first law could be
concluded through the examination
of the catch, without the need of any empirical work; moreover, by the same order of arguments, the second
law may also fail as one cannot fish in all waters.

However, it is clear that there is a fundamental difference between physics and ichthyology.
In the former, it seems
possible to acquire knowledge through purely theoretical epistemological methods. Indeed,
special and general relativity are above all
intellectual constructions. The same can be said about quantum mechanics and, in particular,
about its most fundamental characteristic feature, the Uncertainty Principle.

As already mentioned, modern cosmology is a brain child of general relativity and recent
developments such as inflation and the subsequent imprinting it leaves on the cosmic microwave background
radiation were essentially driven by theoretical problems, both in theoretical cosmology
(the horizon and the flatness problems, and the origin of structure) and in high energy physics
(the overabundance of magnetic
monopoles). The attempt to unify all interactions of nature in a single and encompassing
scheme is a purely theoretical programme. Its most developed proposal,
superstring theory/M-theory does bring about, as any original theoretical construction,
a fairly new view of the Universe. Actually, it seems to suggest a multiverse
(see \cite{BPol}, \cite{Susskind}, \cite{Bertolami06b}, \cite{Bertolami08} for discussions).

However, reality has always the final word and is its quite exciting when surprising or
unexpected possibilities emerge from the observations. The recent discovery of
the current accelerated expansion of the Universe falls precisely in this category.

\section{The Universe as the framework for literature}

For many authors, the Universe, with its laws and dynamics, is an active framework for 
literary expression. Furthermore, attempts to understand how the Universe works are seen by some authors as
guidelines for ethics.  For instance, in the "O Homem Duplicado" (2002) (The Duplicated Man),
the Nobel laureate Portuguese author, Jos\'e
Saramago (1922), asserts the significance of the literary work based on the existence of a cosmic
equilibrium \citep{Saramago}:

\noindent
{\it ``... the conventional tradition of the romance, is not after all, just a somewhat
wasted descriptive attempt due to the scarcity of
imagination, but actually a literary result of the majestic cosmic equilibrium, given that the
universe is, since its origins, a system without any
organizational intelligence, but that had enough time to learn with the infinite multiplication
of its own experiences, so as to abundantly demonstrate
that the performance of life is an infinite machinery of compensation, within each any delay of a
minute, an hour, a century is irrelevant.''}

The failure to assign a clear cut moral sense from descriptions of the origin of the
cosmos is attributed to the lack of consensus around any particular
cosmogony \citep{Saramago}:

\noindent
{\it `` ... It leads one thinking that as all cosmogonies invented since the birth of the word failed so
miserably, it does not mean any good in what concerns their implications for our behaviour.''}

It is interesting to speculate whether Saramago's opinion would change if introduced to
the most recent developments in cosmology and with how observational discoveries
can be harmonized in the context of the Big Bang model. This author suspects that not significantly.

Fernando Pessoa (1888 -- 1935) was the dominant figure of the Portuguese literature in the first
half of the XX century. Multiple literary personas manifest themselves as
``heter\'onimos", Fernando Pessoa, \'Alvaro de Campos, Ricardo Reis, Alberto Caeiro, Bernardo
Soares, Bar\~ao de Teive, Alexander Search, etc. (actually 19 ones) through fairly distinct styles \citep{Pessoa}.
Beyond doubt a unique example in world literature. His work was only partially published during his
lifetime. This rather singular situation has given origin to a great
deal of posthumous publications and quite often to the discovery of unknown poems and sometimes even
whole manuscripts - even though, most often, not completely finished. Rather recently, a new poem by
Alberto Caeiro, the naive and symbolic poet, was found. I present (and translate) one transcription of
this poem, which can be regarded as an ``ode" to the Big Bang:

\noindent
{\it ``I like the sky because I believe it is finite.

\noindent
How could something that has neither a beginning nor an end
have anything to do with me?

\noindent
I do not believe in infinity, I do not believe in eternity.

\noindent
I believe that space starts somewhere and ends somewhere.

\noindent
Beyond and before that there is absolutely nothing.

\noindent
I believe that time has a beginning and an end.

\noindent
Before and after that there was no time.

\noindent
Why any of this should be false? It is false to talk about infinities,

\noindent
As if we know what they are and if we can understand them.

\noindent
No: everything is a finite quantity of things.

\noindent
All is well defined, all has limits, all is made up of things.''}

\section{A cosmic inspired ethics?}
\label{sec:conclusions}

In ancient cultures, given the historical development of a civilization was regarded as
a continuation into the human sphere of a cosmogony which took place in the
natural world. The fact that the latter occurred through a divine intervention would automatically
associate it with a well defined set of religious and ethical values.
Cosmology and religion were once quite intertwined.
This is evident in
the context of the great religions, and this connection can also be found in many other cultures.

Let me illustrate this relationship, through an example based on a passage of the
cosmology of the Mande peoples \citep{Mande},
an ethnic group of West Africa. Speakers of the Mande languages are found in
Gambia, Guinea, Guinea-Bissau, Senegal, Mali, Sierra Leone, Liberia, Burkina Faso, Ivory Coast
and the northern half of Ghana:

\noindent
{\it ``When the Everlasting addressed man, He taught him the law by which all elements of
the Cosmos were formed and continue to exist.
He made man Guardian and Governor of His universe and charged him supervision and
maintenance of universal Harmony.
That is why man is a heavy responsibility.''}

In my view, the key words in this example are {\it universal Harmony} and {\it responsibility}
and I believe that the emphasis on these two concepts is particularly
appealing as they open the possibility of considering a
``cosmic ethics" without relating it to a religious view of the world. If so, the question is if
cosmology can be the cornerstone for an ethics of responsibility. From a strictly scientific point of
view, the answer is clearly negative.
Scientific developments were achieved independently from humanistic and anthropocentric concerns.
The scientific facts that describe and allow for understanding the existence and dynamics of
Earth, home of humankind, are a particular limit of a general set of laws that govern the whole Universe.
It is therefore, somewhat improper to ask for the implication that research on the infinitely large
(and equally well on the infinitely small) might have, on philosophical and ethical terms, for the
future of humankind. Moreover, cosmology does provide, more vociferously than any other subject,
a clear perspective of the modest standing of humankind within the picture of the cosmos.
Nevertheless, cosmology does render us with a
view of how unique, and this is a rather anthropocentric interpretation, are the conditions required to
shelter life and, in particular, sentient and reflective life.
Even though it is a firm belief of this author that life is a wide spread
phenomenon in the Universe, humankind is most likely quite unique within the family of
self-conscious species that exist throughout the Universe. We have therefore, a responsibility to keeping the balance of our
world and ensure its continuity. A responsibility with a time arrow pointing towards
the future, but that is necessarily based on lessons learned from our history, personal and collective.


{\bf Acknowledgments~~}

\noindent
The author is indebted to Ari Belenkiy, Maria da Concei\c c\~ao Bento,
Chanda Carey and Jorge P\'aramos for their constructive comments and suggestions.
This work is partially supported by Funda\c{c}\~ao para a Ci\^encia e a
Tecnologia (Portugal) under the project POCI/FIS/56093/2004.



\bibliographystyle{unstr}

\end{document}